\documentstyle[11pt,newpasp]{article}

\begin{document}

\title{Formation of ellipticals by unequal mass mergers}
\author{Thorsten Naab \& Andreas Burkert}
\affil{Max-Planck-Institut f\"ur Astronomie, K\"onigstuhl 17, 69117
Heidelberg}

\begin{abstract}
Collisionless N-body simulations of merging disk-galaxies with mass
ratios ($\eta$) of 1:1, 2:1, 3:1, and 4:1 have been performed using direct
summation with the special purpose hardware GRAPE. The simulations are
used to examine whether the formation of elliptical galaxies can be
explained in the context of the merger scenario. The photometric,
kinematical and isophotal properties of the merger remnants are
investigated and turn out to be in very good agreement with
observations of giant elliptical galaxies. We conclude that equal mass
mergers lead to slowly rotating, anisotropic remnants having
predominantly boxy isophotes. Mergers with a mass ratio of 3:1 and
4:1, on the other hand, are fast isotropic rotators with disky
isophotes. Projection effects can explain the observed scatter in the
kinematical and isophotal properties of elliptical galaxies.  
\end{abstract}


\section{Merger model and results}
The disk-galaxies are constructed in dynamical equilibrium (Hernquist,
1993) and consist of an exponential stellar disk, a bulge with a
Hernquist profile, and a pseudoisothermal dark halo (units as in
Hernquist, 1993). The two merging galaxies approach each other on
nearly parabolic orbits at a pericenter distance of 2 scale lengths of the
larger disk. The large galaxy is realized with 20000 disk particles,
6666 bulge particles and 40000 halo particles, respectively. The
smaller galaxy contains $(1/\eta)$ of the mass and of the particles in
each component and has a disk scale length of$(1/\eta)^{1/2}$ compared
to the more massive galaxy. We tested 14 different relative
orientations for every mass ratio. The time integration was performed
using the special purpose hardware GRAPE.  
 
After the remnants settled into equilibrium an artificial image of the
remnant was created (see also Heyl, Hernquist \& Spergel,
1994). Following the definition of Bender, D\"oberreiner \&
M\"ollenhoff (1988) we determined the characteristic 
isophotal shape $a4_{\mathrm{eff}}$, ellipticity
$\epsilon_{\mathrm{eff}}$ , the ratio of major axis rotation and
central velocity dispersion, $v_{\mathrm{maj}}/\sigma_0$, and the
anisotropy parameter $(v_{\mathrm{maj}}/\sigma_0)^*$ for 500 random
projections of each of the 14 orbital geometries. These values were
used to calculate a probability density for a given simulated remnant
to be ``observed`` at a given location in the two dimensional
parameter plane, adopting that mergers occur randomly without any
preferred relative inclination. Figure 1 shows the result for 1:1,
2:1, 3:1 and 4:1 merger remnants. From these results we conclude that
most of the global properties of elliptical galaxies can be explained
by a sequence of stellar mergers between disk galaxies of mass ratios
between 1:1 and 4:1. 1:1 mergers completely erase the structure of the
initial disk. In the 3:1 and 4:1 case the remnants seem to remember
their initial state (see Barnes, 1998). In this sense the sequence of
mass ratios is a sequence of disk disruption. However, even with 4:1
remnants we fail to reproduce the fastest observed rotators with
$v_{\mathrm{maj}}/\sigma_0 > 1$ at one effective radius (Figure
1). Recent observations of fast rotating low luminosity ellipticals
(Rix, Carollo \& Freeman, 1999) show that the disagreement is even
stronger at larger radii (see Cretton, Naab, Rix \& Burkert, this
conference)

\begin{figure*}
\includegraphics{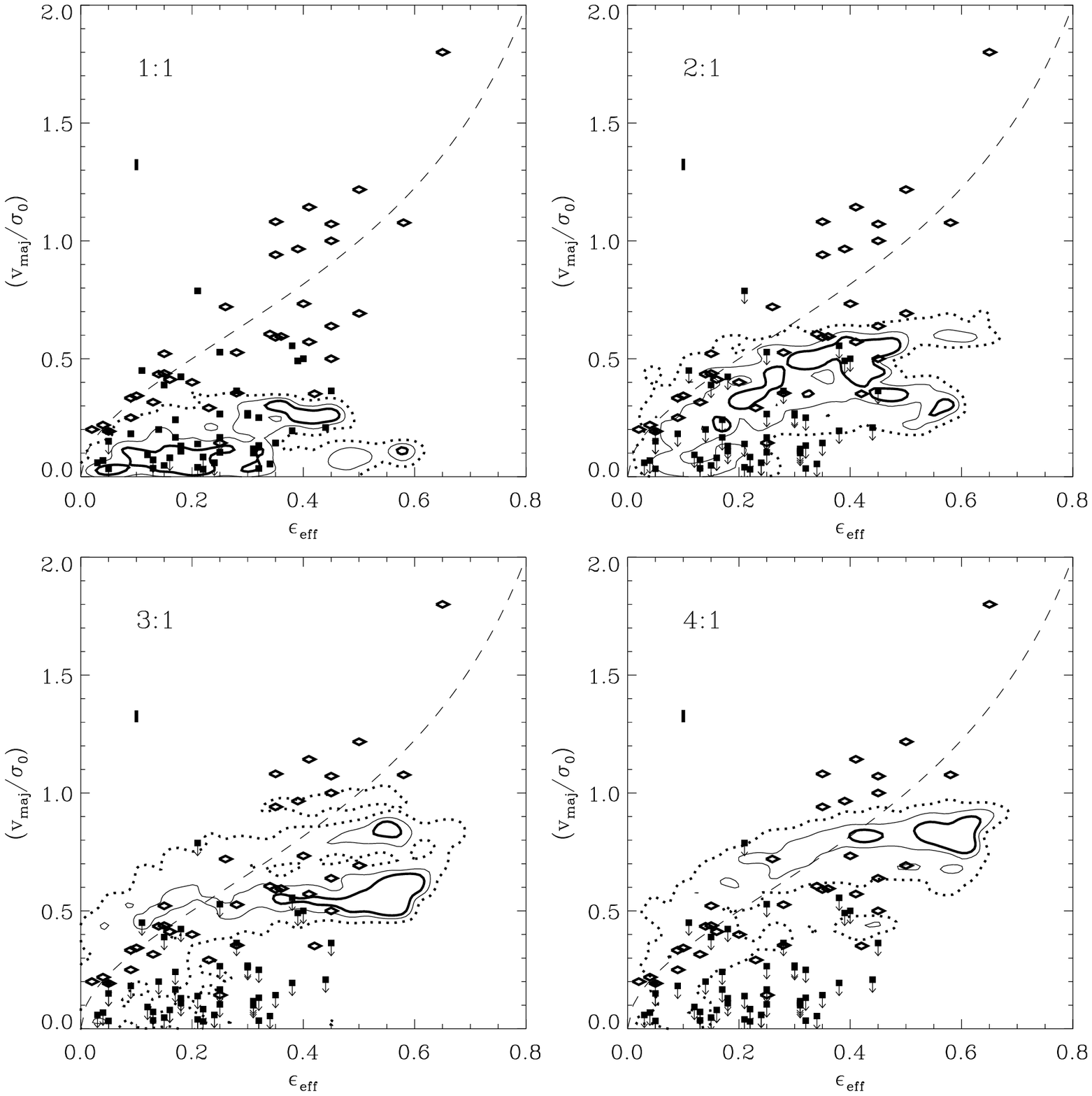}
\includegraphics{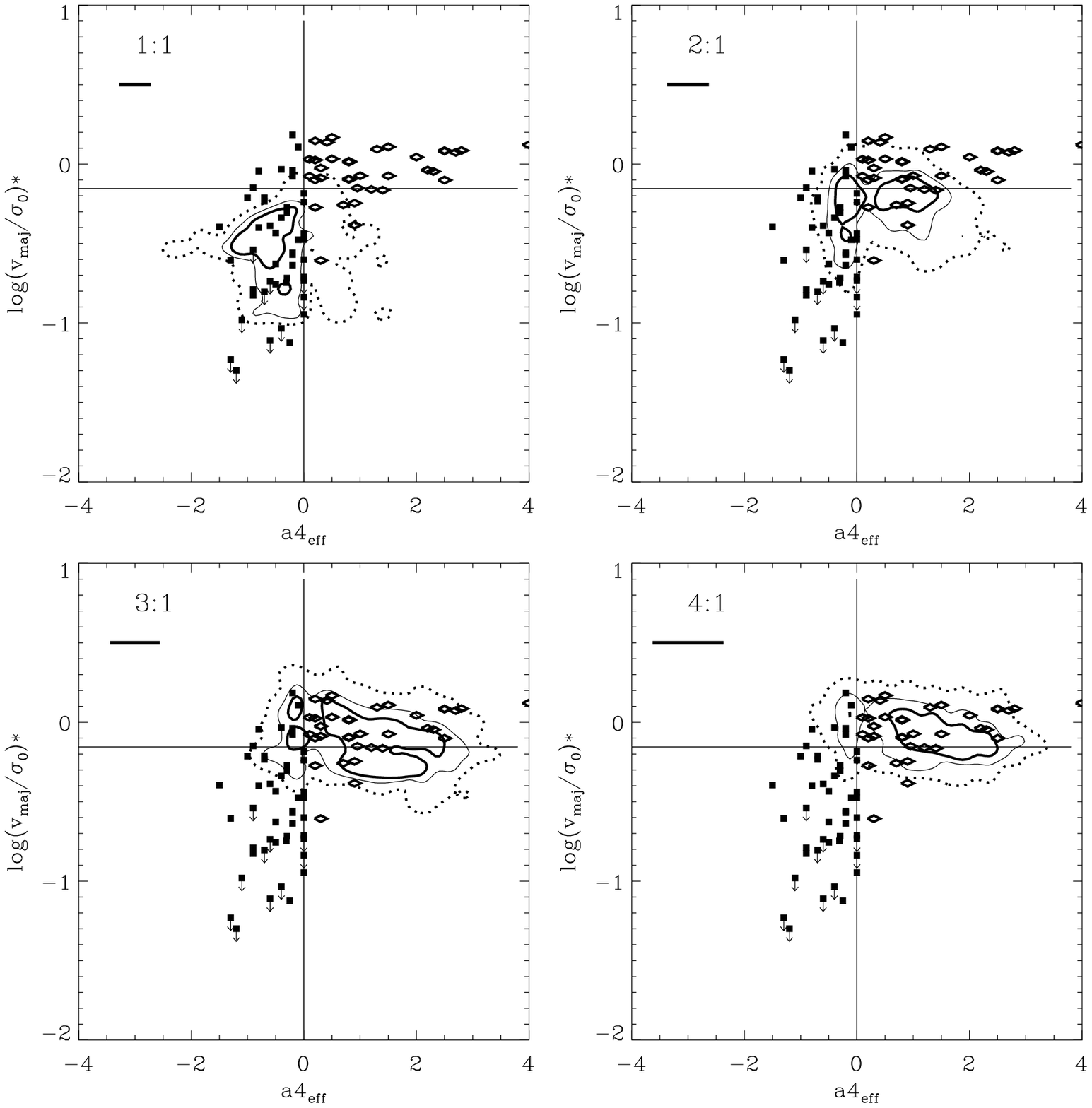}
\vspace{5.8cm}
\caption{{\it Left}:The ratio $v_{\mathrm{maj}}/\sigma_0$ along the major axis
vs. characteristic ellipticity $\epsilon_{\mathrm{eff}}$ for mergers
with a mass ratio of 1:1, 2:1, 3:1, and 4:1. The contours indicate the
50\% (thick line), the 70\% (thin line), and the 90\% (dotted line)
probability to find a merger remnant in the enclosed area. The errors
are calculated by applying statistical bootstrapping. The dashed line
shows the theoretical value for an oblate isotropic rotator. Black
boxes indicate values for observed boxy elliptical galaxies, open
diamonds those for observed disky ellipticals (data kindly provided
by Ralf Bender). {\it Right}: Anisotropy parameter
$(v_{\mathrm{maj}}/\sigma_0)^*$ vs. $a4_{\mathrm{eff}}$.}   
\end{figure*}

\end{document}